\documentclass[twocolumn,appendixfloats]{aastex6}

\begin{document}


\title{Fragmentation of Filamentary Cloud Permeated by Perpendicular Magnetic Field}


\author{Tomoyuki Hanawa}
\affil{Center for Frontier Science, Chiba University\\
1-33 Yayoi-cho, Inage-ku, Chiba\\
Chiba  263-8522, Japan}

\author{Takahiro Kudoh}
\affil{Faculty of Education, Nagasaki University\\ 1-14 Bonkyo-machi, Nagasaki\\
Nagasaki 852-8521, Japan}

\and

\author{Kohji Tomisaka}
\affil{Division of Theoretical Astronomy, National Astronomical Observatory of Japan\\
Mitaka, Tokyo 181-8588, Japan, Department of Astronomical Science, School of Physical
Sciences, SOKENDAI (The Graduate University for Advanced Studies), Mitaka, Tokyo 181-8588, Japan }

\begin{abstract}
We examine the linear stability of an isothermal filamentary cloud permeated 
by a perpendicular magnetic field.   
 Our model cloud is assumed to be supported by gas 
pressure against the self-gravity in the unperturbed state.  For simplicity, the density distribution is 
assumed to be symmetric around the axis.
Also for simplicity, the initial magnetic field is assumed to be uniform and turbulence is not
taken into account.  
The perturbation equation is formulated to be an eigenvalue problem.
The growth rate is obtained as a function of the wavenumber for
fragmentation along the axis and the magnetic field strength.
The growth rate depends critically on the outer boundary.
If the displacement vanishes in the region very far from the cloud
axis (fixed boundary), cloud  fragmentation is suppressed 
by a moderate magnetic field, which means
the plasma beta is below 1.67 on the cloud axis.
If the displacement is constant along the magnetic field in the
region very far from the cloud, the cloud is unstable even when the
magnetic field is infinitely strong.   The cloud  is deformed
by circulation in the plane perpendicular to the magnetic field.
The unstable mode is not likely to induce dynamical collapse,
since it is excited even when the whole cloud is magnetically
subcritical.  For both the boundary conditions the magnetic field increases the 
wavelength  of the most unstable mode.   We find that the
magnetic force suppresses compression perpendicular to
the magnetic field especially in the region of  low density.
\end{abstract}




\section{INTRODUCTION} \label{sec:intro}

It is widely accepted that stars form from the fragmentation  of filamentary clouds
\citep[see, e.g.,][and the references therein]{andre14}.   This idea is supported
by observations showing that the filamentary structure is universal in interstellar
clouds and that prestellar cores and newly formed stars are associated with
the dense parts of filamentary clouds.   It is also well known that filamentary
clouds are unstable against fragmentation if they are gravitationally bound 
\citep[see, e.g.,][and the references therein]{stodolkiewicz63,larson03}.

Fragmentation of a magnetized filamentary cloud has been studied 
extensively \citep[see, e.g.,][]{stodolkiewicz63,nakamura93,hanawa93,fiege00}, particularly
in 1990s.   We note that the magnetic field was generally assumed to be either 
parallel or helical to the cloud axis in these studies.
However, fragmentation depends on the magnetic field direction, since
the magnetic force is perpendicular to the local magnetic field.
We know that massive main filamentary clouds are often associated
with a magnetic field perpendicular to the axis, although less massive
sub-filaments are parallel to the magnetic field \citep[see, e.g.,]
[and the references therein]{sugitani11,palmeirim13,andre14,kusune16}.  
Prestellar cores are often associated with the main
filaments.  Thus, we should examine the effects of the magnetic
field perpendicular to the cloud axis on fragmentation.

It is not easy to construct a self-consistent model of a filamentary molecular
cloud associated with a magnetic field perpendicular to the cloud axis. 
Consequently, it is not easy to analyze the stability against fragmentation.
As shown in \citet{tomisaka14}, the cloud is likely to be flattened, like
fettuccine pasta.  His equilibrium model shows that the cloud is 
supported in part by the magnetic force against gravity.   
The numerically obtained density distribution is difficult to apply in a stability analysis.
\citet{hanawa15} examined fragmentation of a filamentary cloud
permeated by a perpendicular magnetic field under the approximation
that the filamentary cloud is highly flattened in the direction perpendicular
to the magnetic field.   They neglected the density structure
along the magnetic field.

With the abovementioned difficulties in mind, we examine the stability of
a simple self-consistent model for a magnetized filamentary cloud.
The model cloud is assumed to be isothermal and supported 
against the self-gravity solely by the gas pressure in equilibrium.
The initial magnetic field is assumed to be uniform and perpendicular
to the cloud axis so that the magnetic force vanishes in the initial
state.  Although turbulence is dominant over the thermal pressure,
especially in the low density region, we ignore interstellar 
turbulence for simplicity.   Nevertheless, some effects of turbulence may be
taken into account  effectively by replacing the sound speed with the
typical turbulent  velocity.
The model cloud is also assumed to be infinitely long and the density to be 
a function of the radius from the axis.  The magnetic field works
against fragmentation through tension.   We evaluate the effects of the magnetic field
on fragmentation by conducting a linear stability analysis.   The growth
rate of the instability is obtained as a function of the wavelength and
the magnetic field strength, i.e., the initial plasma beta at the cloud center.

This paper is organized as follows.  We describe our equilibrium model
and methods of stability analysis in \S 2.  We employ two types of
boundary conditions, fixed and free,  for the magnetic field in the region
very far from the cloud center.  The result of the stability analysis is 
presented in \S 3.  The growth rate is shown to depend on the choice of boundary
condition. We discuss the implication of our stability analysis in \S 4.
In Appendix A, we prove that  an unstable mode has a real growth rate.
In Appendix B, we describe the elements of the matrixes used for
our stability analysis.
In Appendix C, we calculate the growth rate for an unmagnetized filamentary
cloud in order to evaluate the accuracy of our numerical analysis.

\section{METHODS}

\subsection{Basic Equations}

For our stability analysis, we employ the ideal magnetohydrodymamic (MHD) equations, 
\begin{eqnarray}
\frac{\partial \rho}{\partial t} + \mbox{\boldmath$\nabla$} \cdot
\left( \rho \mbox{\boldmath$v$} \right) = 0 , \\
\rho \frac{d \mbox{\boldmath$v$}}{dt} = - \mbox{\boldmath$\nabla$} P 
+ \mbox{\boldmath$j$} \times \mbox{\boldmath$B$} + 
\rho \mbox{\boldmath$\nabla$} \Phi ,  \label{motion_eq} \\
\mbox{\boldmath$j$} = \frac{\mbox{\boldmath$\nabla$} \times \mbox{\boldmath$B$}}{4 \pi} , 
\label{e-current}
\end{eqnarray}
where $ \rho $, $ P $, $ \Phi$ $, $ \mbox{\boldmath$v$}$, $ \mbox{\boldmath$B$} $, 
and $ \mbox{\boldmath$j$} denote the density, pressure, gravitational potential, velocity,
magnetic field, and electric current density, respectively.   The gas is assumed to be isothermal and so
the equation of state is expressed as
\begin{eqnarray}
P & = & \rho c _s ^2 ,
\end{eqnarray}
where $ c _s $ denotes the sound speed. 
 The self-gravity of the gas is taken into account 
through the Poisson equation 
\begin{eqnarray}
\Delta \Phi = 4 \pi G \rho ,
\end{eqnarray}
where $ G $ denotes the gravitational constant.  We ignore ambipolar diffusion and 
turbulence for simplicity. 

\subsection{Equilibrium Model}

We consider an isothermal filamentary cloud in hydrostatic equilibrium.
The cloud axis is assumed to be located on $ x = y = 0 $ in Cartesian
coordinates.  Then the density distribution is expressed as
\begin{eqnarray}
\rho _0  (x, y, z) & = & \rho _c \left( 1 + \frac{x ^2 + y ^2}{8 H^2} \right) ^{-2} , \\
H ^2 & = & \frac{c _s ^2}{4 \pi G \rho _c} , \label{Hrhoc}
\end{eqnarray}
where $ \rho _0 $ and $ \rho _c $ denote the density distribution in equilibrium and the value on the cloud
axis, respectively \citep{stodolkiewicz63,ostriker64}.  

 We assume that the cloud in equilibrium is permeated by a uniform magnetic field in the $ x $-direction,
\begin{equation}
\mbox{\boldmath{$B$}} _0  =  B _0 \mbox{\boldmath$e$} _x ,
\end{equation}
where $ \mbox{\boldmath$e$} _x $ denotes the unit vector in the $ x $-direction.
This uniform magnetic field does affect the cloud stability but not the equilibrium.

To specify the initial magnetic field strength in the analysis, we use the plasma beta at the cloud center,
\begin{eqnarray}
\beta & = & \frac{8 \pi \rho _c c _s ^2}{B _0 ^2} .
\end{eqnarray}
The plasma beta is related to the
mass to flux ratio,
\begin{eqnarray}
f & = & \frac{\displaystyle \int \rho _0 (x ^\prime, y, z) dx ^\prime}{B _0} \\
& = & \frac{\sqrt{2} \pi \rho _c H}{B _0} \left(1 + \frac{y ^2}{8 H ^2} \right) ^{-3/2} .
\end{eqnarray}
The mass to flux ratio can be rewritten as 
\begin{eqnarray}
f  & = & \frac{1}{4} \sqrt{\displaystyle \frac{\beta}{G}} 
\left( 1 + \frac{y^2}{8H ^2} \right) ^{-3/2} ,
\end{eqnarray}
by using Equation (\ref{Hrhoc}).  The critical mass to flux ratio is
$ f _c = 1/(2 \pi \sqrt{G}) $.   Thus, the cloud is subcritical when
\begin{eqnarray}
\beta & < & 0.405 \left(1 + \frac{y^2}{8H^2} \right) ^3. \label{betac}
\end{eqnarray}

\subsection{Perturbation Equation}

We consider a small perturbation around the equilibrium in order to search for an unstable
mode. The perturbation is described by the displacement defined by  
\begin{eqnarray}
\mbox{\boldmath$\xi$} & = & \xi _{x} (x, y) \cos k z \mbox{\boldmath${e}$} _x +
\xi _y (x, y) \cos k z \mbox{\boldmath$e$} _y 
\nonumber \\
& & + \xi  _z (x, y)  \sin k z \mbox{\boldmath$e$} _z , \label{displacement} 
\end{eqnarray}
where the perturbation is assumed to be sinusoidal in the $ z $-direction with the
wavenumber $ k $.   
The density perturbation is derived from the equation of continuity,
\begin{equation}
\delta \rho   =  - \mbox{\boldmath$\nabla$} \cdot 
\left( \rho _0 \mbox{\boldmath$\xi$} \right) .  \label{continuity}
\end{equation}
Substituting Equation (\ref{displacement}) into Equation (\ref{continuity}), we obtain
\begin{eqnarray}
\delta \rho (x, y, z) & = & \delta \varrho (x, y) \cos k z , \\
\delta \varrho & = & - \frac{\partial}{\partial x} \left( \rho _0 \xi _x \right) -
\frac{\partial}{\partial y} \left( \rho _0 \xi _y \right) - k \rho _0 \xi _z . 
\label{deltarho}
\end{eqnarray}
Similarly, we obtain the perturbation in the magnetic field from the induction
equation,
\begin{equation}
\delta \mbox{\boldmath$B$}  = \mbox{\boldmath$\nabla$} \times 
\left( \mbox{\boldmath$\xi$} \times \mbox{\boldmath$B$} _0 \right) .
\end{equation}
The induction equation  is further expressed as
\begin{eqnarray}
\delta \mbox{\boldmath$B$} (x, y, z) & = & b _x (x, y) \cos k z \mbox{\boldmath$e$}_x
 +  b _y (x, y) \cos k z \mbox{\boldmath$e$} _y \nonumber  \\
& \; &  + b _z (x, y) \sin k z \mbox{\boldmath$e$} _z  , \\
b _x & = & - B _0 \left[  \frac{\partial}{\partial y} \xi _y (x, y) + k \xi _z \right] , \label{defbx}\\
b _y & = & B _0 \frac{\partial \xi _y}{\partial x} , \label{defby} \\
b _z & = & B _0 \frac{\partial \xi _z}{\partial x} . \label{defbz}
\end{eqnarray}
We evaluate the change in the current density to be
\begin{equation}
\delta \mbox{\boldmath$J$}  =  \frac{1}{4\pi} \mbox{\boldmath$\nabla$}
\times \delta \mbox{\boldmath$B$} ,
\end{equation}
using Equation (\ref{e-current}).  Each component of the current density is expressed as
\begin{eqnarray}
\delta \mbox{\boldmath$J$} (x, y, z)  & = & j _x (x,y) \sin k z \mbox{\boldmath$e$} _x +
 j _y (x,y) \sin k z \mbox{\boldmath$e$} _x \nonumber \\
 & \; &  + j _z (x, y) \cos k z \mbox{\boldmath$e$} _z , \\
 j _x  & = & \frac{1}{4 \pi} \left( \frac{\partial b _z}{\partial y} + k b _y \right) ,  \\
 j _y  & = & - \frac{1}{4 \pi} \left( k  b _x + \frac{\partial b _z}{\partial x} \right) , \\
 j _z  & = & \frac{1}{4 \pi} \left( \frac{\partial b _y}{\partial x} 
 - \frac{\partial b _x}{\partial y} \right) .
\end{eqnarray}
Then the changes in the density and current density are expressed as an explicit function of
$ \mbox{\boldmath$\xi$} $.

The change in the gravitational potential is given as the solution of the Poisson equation
\begin{equation}
\mbox{\boldmath$\nabla$} ^2 \delta \psi = 4 \pi G \delta \rho . \label{poisson}
\end{equation}
Thus, it can be regarded as an implicit function of $ \mbox{\boldmath$\xi$} $.

We derive the equation of motion for the perturbation by taking account of the
force balance,
\begin{equation}
\mbox{\boldmath$\nabla$} \left( c _s ^2 \rho _0 \right) + \rho _0 
\mbox{\boldmath$\nabla$} \psi _0 = 0, 
\end{equation}
with no electric current density, $ \mbox{\boldmath$j$} _0 = 0 $,
in equilibrium.
Then the equation of motion is expressed as
\begin{equation}
\sigma ^2 \rho _0 \mbox{\boldmath$\xi$} = - c _s ^2 \rho _0 
\mbox{\boldmath$\nabla$} \left( \frac{\delta \rho}{\rho _0} \right)
 -  \rho _0 \nabla \delta \psi 
+ \delta \mbox{\boldmath$J$} \times \mbox{\boldmath$B$} _0 , \label{motion}
\end{equation}
where the last term represents the magnetic force.  The linear growth rate, $ \sigma $,
is obtained as the eigenvalue of the differential equation (\ref{motion}), since the
right-hand side  is proportional to $ \mbox{\boldmath$\xi$} $.
As shown in Appendix A, the growth rate should be either real or pure imaginary.

\begin{table}[h]
\caption{Variables Describing Perturbations\label{symmetry}}
\begin{center}
\begin{tabular}{llcc}
\hline
variable  & evaluation  & symmetry &  symmetry \\
& point &  $ x $ & $ y $ \\
\hline
$ \xi _x  $ & $ (i-1/2,j) $ & A & S\\
$ \xi _y  $ & $ (i,j-1/2) $  & S & A \\
$ \xi _z  $ & $ (i,j) $ &   S & S \\
$ \delta \varrho$ & $ (i,j) $ &   S & S \\
$ \delta \psi $ & $ (i,j) $ &   S & S \\
$ b_x  $ & $ (i,j) $ &   S & S \\
$ b _y  $ & $ (i-1/2,j-1/2) $ & A & A \\
$ b _z  $ & $ (i-1/2,j) $  & A & S \\
$ j _y  $ & $ (i,j) $ &  S & S \\
$ j _z  $ & $ (i,j-1/2) $ & S & A  \\
\hline
\end{tabular}
\end{center}
\tablecomments{A: anti-symmetric. S: 
symmetric.}
\end{table}

The equilibrium model is symmetric with respect to the $ x $- and $ y $-axes.   Thus, all 
eigenmodes should be either symmetric or anti-symmetric with respect to these axes.
We restrict ourselves to the eigenmodes symmetric to both $ x $- and 
$ y $-axes, since the unstable mode has the same symmetry in the case of
no magnetic field \citep{nakamura93}.  The choice of this symmetry is justified since
we are interested only in the unstable mode.    Using this symmetry, we can reduce the
region of computation to the first quadrant, $ x \ge 0 $ and $ y \ge 0 $.
The variables describing the perturbation and their symmetries are summarized in Table 
\ref{symmetry}. 

We consider two types of the boundary conditions.   The first one assumes that  
the displacement should vanish in the region very far from the filament center.  
We call this the fixed boundary since the magnetic field lines are fixed on the boundary.  
The second one allows the magnetic field lines to move while remaining straight and 
normal to the boundary.   This restriction is expressed as 
\begin{eqnarray}
\left( \mbox{\boldmath$B$} _0 \cdot \mbox{\boldmath$\nabla$} \right) \mbox{\boldmath$\xi$} 
= 0 .
\end{eqnarray}
Thus, we assume $ \partial \mbox{\boldmath$\xi$} / \partial x $ on the boundary in
the $ x $-direction and $ \mbox{\boldmath$\xi$} = 0 $ in the $ y $-direction.
We refer to this as the free boundary condition.   In both types of boundary conditions, we use the symmetries
given in Table \ref{symmetry} 
to set the boundary conditions for $ x = 0 $ and $ y = 0 $.

\subsection{Numerical Methods}

We solve the eigenvalue problem numerically by a finite difference approach.  The differential
equations are evaluated on the rectangular grid in the $xy$ plane.   We evaluate $ \xi _z $,
$ \delta \varrho $, $ \delta \psi $, $ b _x $,  and $ j _y $ at the points
\begin{eqnarray}
\left( x _i, y _j \right) & = & \left( i \Delta x, j \Delta y \right) ,
\end{eqnarray}
where $ i $ and $ j $ specify the grid points, while $ \Delta x $ and $ \Delta y $ denote 
the grid spacing in the $ x $- and $ y $-directions, respectively (see Table \ref{symmetry}).   
These variables are symmetric with respect to both the $ x $- and $ y $-axes.   Using this
symmetry, we consider the range $ 0 \le i \le n _x $ and $ 0 \le j \le  n _y $, where
$ n _x $ and $ n _y $ specify the number of grid points in each direction.  
When $ i > n _x $ or $ j > n _y $, the displacement
$ \xi _{z,i,j} $ is assumed to vanish  for
the fixed boundary and to have the same values at neighboring points
in the computation domain for the free boundary condition.    
We use the indexes,
$ i $ and $ j $, to specify the position where the variables are evaluated, such
as $ \xi _{z,i,j} = \xi _z (x _i, y _j) $.  

The variables $ \xi _x $ and $ b _z $ are evaluated at the points
\begin{eqnarray}
\left( x _{i-1/2}, y _j \right) & = & \left[ \left( i - \frac{1}{2} \right)  \Delta x, j \Delta y \right] .
\end{eqnarray}
These variables are anti-symmetric with respect to $ x $ and symmetric with
respect to $ y $.   Similarly, the variables $ \xi _y $ and $ j _z $ are evaluated at
the points 
\begin{eqnarray}
\left( x _i, y _{j-1/2} \right) & = & \left[ i \Delta x, \left( j - \frac{1}{2} \right) \Delta y \right] ,
\end{eqnarray}
since they are symmetric with respect to $ x $ and anti-symmetric with respect to $ y $.
Given that $ b _y $ is anti-symmetric with respect to both $ x $ and $ y $, it is evaluated at
the points
\begin{eqnarray}
\left( x _{i-1/2}, y _{j-1/2} \right) & = & \left[\left( i - \frac{1}{2} \right) \Delta x,
 \left( j - \frac{1}{2} \right) \Delta y \right] .
\end{eqnarray}
All these variables are evaluated in the region $ 0 \le x \le n _x \Delta x $ and
$ 0 \le y \le n _y \Delta y $.
In other words, we use staggered grids to achieve second-order accuracy in space.

Using the variables defined on the grids, we rewrite the perturbation equations.
Equation (\ref{deltarho}) is rewritten as
\begin{eqnarray}
\delta \varrho _{i,j} & = &  - \frac{\rho _{0,i+1/2,j} \xi _{x,i+1/2,j} - \rho _{0,i-1/2,j} \xi _{x,i-1/2,j}}{\Delta x} 
\nonumber \\
& - &   \frac{\rho _{0,i,j+1/2} \xi _{y,i,j+1/2} - \rho _{0,i,j-1/2} \xi _{y,i,j-1/2}}{\Delta y} 
\nonumber \\
& - & k \rho _{0,i,j} \xi _{z,i,j} . \label{continuity3}
\end{eqnarray}
Equation (\ref{poisson}), the Poisson equation, is expressed as
\begin{eqnarray}
\frac{\delta \psi _{i+1,j} + \delta \psi _{i-1,j}}{\Delta x ^2}  
+ \frac{\delta \psi _{i,j+1} + \delta \psi _{i,j-1}}{\Delta y ^2}  \nonumber \\
- \left( \frac{2}{\Delta x ^2} + \frac{2}{\Delta y ^2} +  k ^2 \right) \delta \psi _{j,k} 
 = 4 \pi G \delta \varrho _{i,j} . \label{poisson2}
\end{eqnarray}
The solution of Equation (\ref{poisson2}) is expressed as
\begin{eqnarray}
\delta \psi _{i,j} & = & \sum _{i^\prime} \sum _{j^\prime}
G _{i,j,i^\prime, j^\prime} \delta \varrho _{i^\prime,j^\prime} , \label{poisson3}
\end{eqnarray}
where $ G _{i,j,i^\prime,j^\prime} $ denotes the Green's function and
the value is obtained by solving Equation (\ref{poisson2}) numerically.
The change in the magnetic field  is
evaluated as
\begin{eqnarray}
b _{x,i,j} & = & - B _0 \left( \frac{\xi _{y,i,j+1/2} - \xi _{y,i,j-1/2}}{\Delta y}
+ k \xi _{z,i,j} \right) , 
\end{eqnarray}
\begin{eqnarray}
b _{y,i-1/2,j-1/2} & = & B _0 \left( \frac{\xi _{y,i,j-1/2} - \xi _{y,i-1,j-1/2}}{\Delta x} \right) , \\
b _{z,i-1/2,j} & = & B _0 \left( \frac{\xi _{z,i,j} - \xi _{z,i-1,j}}{\Delta x} \right) ,
\end{eqnarray}
from Equations (\ref{defbx}) through
(\ref{defbz}).
The current density is evaluated as
\begin{eqnarray}
j _{y,i,j} & = & - \frac{1}{4 \pi} \left( k b _{x,i,j} + \frac{b _{z,i+1/2,j} - b _{z,i-1/2,j}}{\Delta x} \right) , 
\end{eqnarray}
\begin{eqnarray}
j _{z,i,j-1/2} & = & \frac{1}{4 \pi} \left( \frac{b _{y,i+1/2,j-1/2} - b _{y,i-1/2,j-1/2}}{\Delta x} \right.
\nonumber \\
& \; & \left.  -  \frac{b _{x,i,j} - b _{x,i,j-1}}{\Delta y} \right) . \label{jz}
\end{eqnarray}
The $ x $-component of the current density, $ j _x $, is not evaluated, since it does not appear in the
equation of motion.
The fixed boundary conditions are expressed as
\begin{eqnarray}
\xi _{x,n_x +1/2,j} & = & 0 , \\
\xi _{y,n_x +1, j-1/2} & = & 0 , \\
\xi _{z,n_x+1,j} & = & 0 , \\
\xi _{x,i-1/2, n_y+1} & = & 0 , \\
\xi _{y,i,n_y+1/2 } & = & 0 , \\
\xi _{z,i,n_y+1} & = & 0 .
\end{eqnarray}
When the free boundary is applied, the conditions are replaced with
\begin{eqnarray}
\xi _{x,n_x+1/2,j} & = & \xi _{x,n_x-1/2,j} , \\
\xi _{y,n_x+1,j-1/2} & = & \xi _{y,n_x,j-1/2} , \\
\xi _{z,n_x+1,j} & = & \xi _{z,n_x,j} , \\
\xi _{x,i-1/2, n_y+1} & = & \xi _{x,i-1/2, n_y} , \\
\xi _{y,i,n_y+1/2 } & = & \xi _{y,i,n_y-1/2} , \\
\xi _{z,i,n_y+1} & = & \xi _{z,i,n_y} .
\end{eqnarray}

The equation of motion (\ref{motion}) is expressed as
\begin{eqnarray}
\sigma ^2 \rho _{0,i-1/2,j} \xi _{x,i-1/2,j} & = & - \frac{c _s ^2\rho _{0,i-1/2,j}}{\Delta x} 
\left( \frac{\delta \varrho _{i,j}}{\rho _{0,i,j}} - \frac{\delta \varrho _{i-1,j}}{\rho _{0,i-1,j}} \right) \nonumber \\
& \; & - \frac{\rho _{0,i-1/2,j}}{\Delta x} \left( \delta \psi _{i,j} - \delta \psi _{i-1,j} \right) . \label{motion1} \\
\sigma ^2 \rho _{0,i,j-1/2} \xi _{y,i,j-1/2} & = & 
- \frac{c _s ^2 \rho _{0,i,j-1/2}}{\Delta y} 
\left( \frac{\delta \varrho _{i,j}}{\rho _{0,i,j}} - \frac{\delta \varrho _{i,j-1}}{\rho _{0,i,j-1}} \right) \nonumber \\
& \; & - \frac{\rho _{0,i,j-1/2}}{\Delta y} \left( \delta \psi _{i,j} - \delta \psi _{i,j-1} \right) \nonumber \\
& \; & + B _0 j _{z,i,j-1/2}  . \label{motion2} \\
\sigma ^2 \rho _{0,i,j} \xi _{z,i,j} & = & - k c _s ^2 \delta \varrho _{i,j} - k \rho _{0,i,j} \delta \psi _{i,j} 
\nonumber \\
& \; &  - B _0 j _{y,i,j} . \label{motion3}
\end{eqnarray}

Equations (\ref{motion1}) through (\ref{motion3}) are summarized in the form
\begin{eqnarray}
\sigma ^2 \mbox{\boldmath$B$} \mbox{\boldmath$\zeta$} = \left( \mbox{\boldmath$A$} + B _0 ^2
\mbox{\boldmath$C$} \right) \mbox{\boldmath$\zeta$} \label{algebraic}
\end{eqnarray}
by using Equations (\ref{continuity3}), and (\ref{poisson3})  through (\ref{jz}).  Here,
$ \mbox{\boldmath$\zeta$} $ denotes an array of components, 
$ \xi _{x,i-1/2,j} $, $ \xi _{y,i,j-1/2} $, and $ \xi _{z,i,j} $ for all the combinations of $ i $ and $ j $.
The matrix elements of $ \mbox{\boldmath$A$} $, $ \mbox{\boldmath$B$} $, and
$ \mbox{\boldmath$C$} $ are evaluated numerically as a function of $ k $.   
See Appendix \ref{matrix-elements} for further details.
Then the growth rate is given as the solution of 
\begin{eqnarray}
\det \left(  \sigma ^2 \mbox{\boldmath$B$} - \mbox{\boldmath$A$} - B _0 ^2 
\mbox{\boldmath$C$} \right) = 0 . \label{generalE}
\end{eqnarray}
We use the subroutine DGGEVX of LAPACK 
\citep[see,][for the software]{anderson99} to solve Equation (\ref{generalE}).
The subroutine returns all the eigenvalues $ \sigma ^2 $.     

The matrixes $ \mbox{\boldmath$A$} $, $ \mbox{\boldmath$B$} $, and
$ \mbox{\boldmath$C$} $ have 
dimension $ \left( 3 n _x n _y + 2 n _x + 2 n _y + 1 \right) $. Thus, we obtain
$ 3 n _x n _y + 2 n _x + 2 n _y + 1 $ eigenmodes.  However, we select only
one unstable mode ($ \sigma ^2 > 0 $) for a given $ k $ and $ B _0 $.
The remaining eigenmodes denote oscillation of the filamentary cloud.
In the following, we restrict ourselves to the unstable mode.

\section{RESULTS}

\subsection{Growth Rate}

\begin{figure}
\plotone{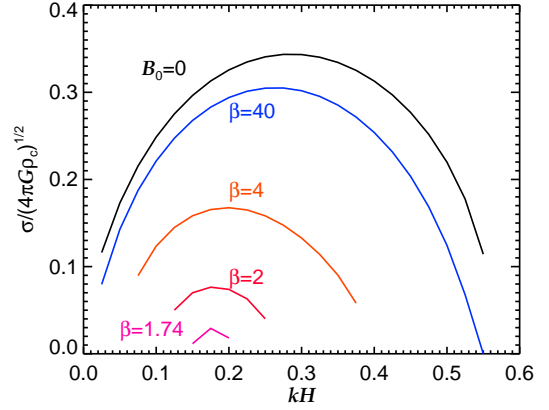}
\caption{Growth rate $ \sigma  $ as a function of the wavenumber $ k H$ for the
fixed boundary.   The ordinate is normalized to units of $ \sqrt{4 \pi G \rho _c} $. \label{growth}}
\end{figure}

Figure \ref{growth} shows the eigenvalue $\sigma$ for the fixed boundary
condition as a function of $ k H $. 
Each curve denotes $\sigma$ in units of 
$\sqrt{4\pi G \rho _c}$ for a given $ B _0 $.
The grid spacing is
set as $ \Delta x = \Delta y = 0.6 H $ and the computation region is
specified as $ n _x = n _y = 40 $.  Thus, the outer boundary is set as
$ x = 24 H $ and $ y = 24 H $.  The growth rate is evaluated over 
the interval $ \Delta( k  H) = 2.5 \times 10 ^{-2} $.
Figure \ref{growth} shows the eigenvalue only when $ \sigma > 3.16 \times 10 ^{-3} 
\sqrt{4\pi G \rho _c} $, i.e.,
 $ \sigma ^2  > 10 ^{-5}~4\pi G \rho _c $.
The imaginary part of $ \sigma ^2 $ is omitted since it is negligibly small.  We 
find at most one growing mode for a given pair of $ k $ and $ B _0 $.

The growth rate $ \sigma $ has its maximum around $ k H \simeq 0.29 $ in the absence of the
magnetic field ($ B _0 = 0 $).   The growth rate is reduced greatly by the magnetic field
having its root fixed at infinity.  We find an unstable mode only in the
range $ 0.05 < kH < 0.4 $ for $ \beta = 4 $.
The unstable mode disappears when $ B _0 > 1.1  
\sqrt{4 \pi \rho _c c _s ^2 } $, i.e., when the plasma beta is smaller than 1.67 ($ \beta < 1.67 $).

\begin{figure}
\plotone{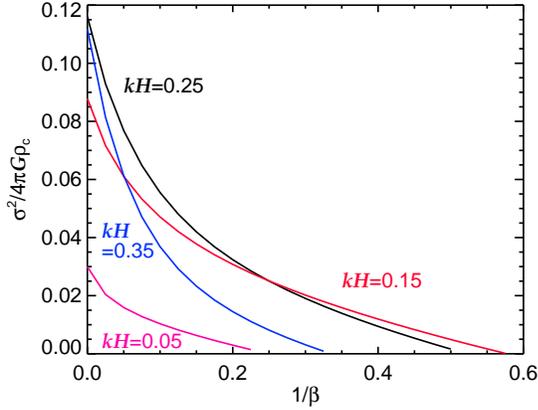}
\caption{Growth squared as a function of $ 1/\beta $ for a given $ kH $, 
where $ \beta $ denotes the initial plasma beta at the cloud center, $ x = y = 0 $. \label{growth2}}
\end{figure}

The stabilization due to the magnetic field is shown clearly in Figure \ref{growth2},
where each curve is $ \sigma ^2 $ as a function of 
$ 1/\beta = B _0 ^2 / (4 \pi \rho _c c _s ^2) $ for a given $ k H $.   
The square of the growth rate, $ \sigma ^2 $, decreases in  proportion to $ 1/ \beta$, 
i.e., $ B _0 ^2 $ in the range $ \beta < 10 $.   The proportional constant is larger for a larger 
$ k H$.  The dispersion relation is similar to that for the MHD fast wave, 
\begin{equation}
\omega ^2  =  \left( c _s ^2 + \frac{\mbox{\boldmath$B$} _0  ^2}{4 \pi \rho _0} \right) k ^2 ,
\end{equation}
where the wave is assumed to propagate normal to the magnetic field $ \mbox{\boldmath$B$} $.

\begin{figure}
\plotone{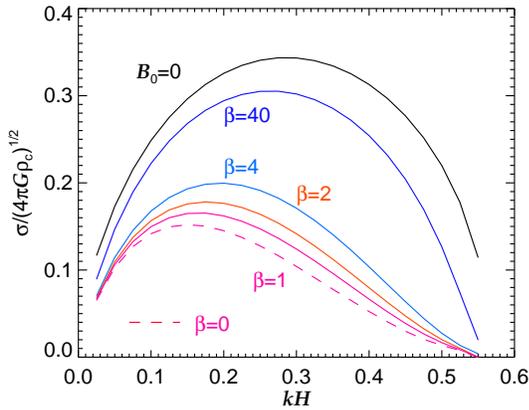}
\caption{Same as Fig. \ref{growth} except for the free boundary. \label{growthML}}
\end{figure}

The growth rate shows a different dependence on the magnetic field strength when 
we apply the free boundary
condition.  Figure \ref{growthML} is the same as Figure \ref{growth} except for the free boundary.
Each curve in Figure \ref{growthML} denotes the growth rate as a function of $ kH $ for a given $ \beta $
whose value is designated by the same color.  The dashed line denotes the
growth rate when $ \beta = 0 $, and the value is obtained by extrapolation.
As the magnetic field strength increases, the growth rate decreases but remains positive for 
$ k H \le 0.525 $, and therefore the cloud is unstable even when the magnetic field is infinitely strong.

The result shown in Figure \ref{growthML} seems to contradict the naive expectation that a subcritical cloud is stable
 \citep[see, e.g.,][]{nakano78}. 
 We explain the apparent discrepancy in the next
subsection. 

\begin{figure}
\plotone{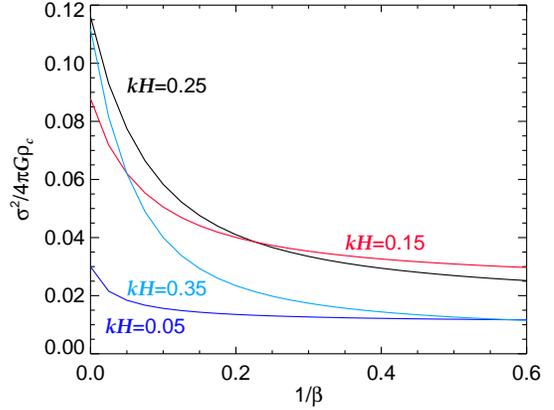}
\caption{Growth squared as a function of $ 1/\beta $ for a given $ kH $ for
the free boundary.  \label{growthMF2}}
\end{figure}

Figure \ref{growthMF2} is the same as Figure \ref{growth2} except for the free boundary.
The growth rate approaches a certain value in the limit of large $ B _0 $.  The growth rate is well
approximated by
\begin{eqnarray}
\sigma ^2 \left( k, B _0 \right) & = & [ \sigma_\infty  (k) ] ^2  +
\left[ \sigma ^\prime (k) \right] ^2 \beta .
\end{eqnarray}
The growth rate $ \sigma _\infty (k) $ is evaluated by a spline fit to the growth rate
in the range $ 0.2 \le \beta \le 1 $.  The value of $ \sigma _\infty (k) $ is shown by
the dashed curve in Figure \ref{growthM}.

When $ kH $ is smaller, the growth rate is close to the asymptotic value for a weaker
magnetic field.   This result implies that the magnetic field is bent or compressed little by
the unstable perturbation  when the initial magnetic field is strong.  Otherwise, the
distorted magnetic field  would induce a strong force acting on the gas.  At the same time,
the displacement should be appreciable even in the region very far from the cloud
center.  Otherwise, the growth rate would be independent of the boundary condition
in the $ x $-direction.   We confirm this expectation in the next subsection.

The wavenumber of the most unstable mode is smaller for a larger $ B _0 $.
The wavenumber is $ kH  \simeq 0.29 $ for $ B _0 = 0 $ and $  k H \simeq 0.16 $ for a small $ \beta $. 
The magnetic field decreases the growth rate and increases the wavelength of a typical
perturbation.  

The filamentary cloud is stable for any perturbation having a wavenumber larger than
$ k > k _{\rm cr} = 0.565 \, H ^{-1} $ when $ B _0 = 0 $.   
When the free boundary is applied, the critical wavenumber is slightly reduced, i.e., 
$ k _{\rm cr} \, H ^{-1} \simeq 0.51 $ for a large $ B _0 $.   The critical wavenumber is
evaluated from the high resolution computation of $ \Delta x = \Delta y = 0.3 H $ and
$ n _x = n _y = 80 $.

\subsection{Eigenmode}

\begin{figure}
\plotone{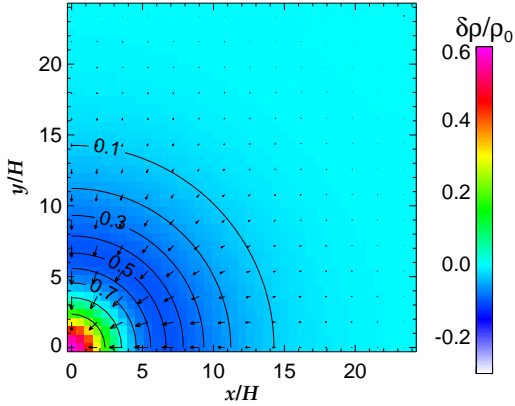}
\caption{Eigenmode for $ B _0 = 0 $ and $ kH = 0.2$. The color
denotes $ \delta \varrho / \rho _0 $, while the contours are of $ - \xi _z/H $.  The arrows
indicate $ (\xi _x, \xi _y) $. \label{eigenB0} }
\end{figure}

First, based on our stability analysis, we review the unstable perturbation in the absence of the magnetic field.
When $ B _0 = 0 $, both the growth rate
and the eigenmode depend little on the boundary condition.  Figure \ref{eigenB0}
shows the unstable mode for the fixed boundary when $ B _0 = 0 $ and $ kH = 0.2 $.
The values are normalized so that $ \xi _z = - H $ at the origin.  The arrows indicate the displacement
$ (\xi _x, \xi _y ) $, while  the contours are of $ \xi _z $.
The density perturbation $ \delta \varrho / \rho _0 $  is representing according to the color scale
given in the right panel.   The displacement is very
small in the region $ \sqrt{x ^2 + y ^2} > 15 H $ when the fixed boundary
is set as $ x = 24 H $ and $ y = 24 H $.   Thus, the growth rate depends little
on the outer boundary condition.   The gas concentrates towards the
origin and the density decreases in the surrounding gas. 

\begin{figure}
\plotone{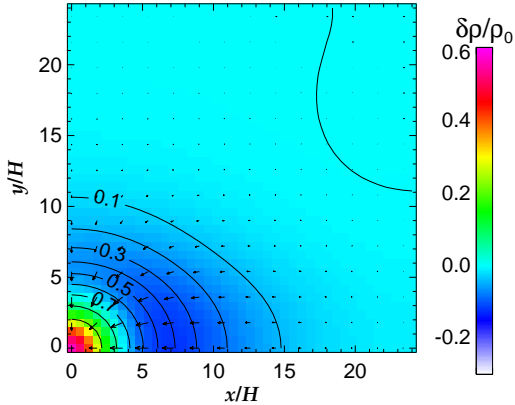}
\caption{Eigenmode for the fixed boundary.  The wavenumber
and the initial plasma beta are set as $ k H = 0.2 $ and $ \beta = 100 $,
respectively.   The notation
is the same as that of Fig.~\ref{eigenB0}. \label{eigenB002} }
\end{figure}

\begin{figure}
\plotone{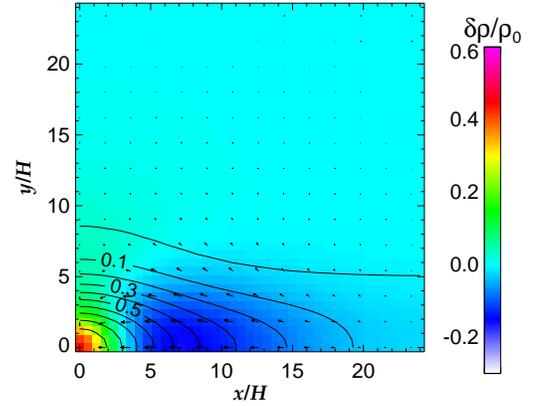}
\caption{Same as Fig.~\ref{eigenB002} except $ \beta = 16.67 $. 
 \label{eigenB005} }
\end{figure}

Low-density gas is affected by a magnetic field even when the magnetic field is very weak.
Figures \ref{eigenB002} and \ref{eigenB005} are the same as Figure
\ref{eigenB0}, except the initial plasma betas are
$ \beta = 100 $ and 16.67 at the cloud center, respectively.
The displacement in the $ y $-direction is greatly suppressed, especially
in the region of large $ y $, since it is perpendicular to the magnetic field.
Also, the displacement along the cloud axis ($ \xi _z$: contour lines) is restricted in the region
$ |y| < 8H $ when $ \beta  = $ 16.67 .   It should be noted that $ | \xi _x | $ is 
large in the more extended region in the $x $-direction (along the initial
magnetic field) than in the absence of the magnetic field.  The gas in the region
far from the cloud axis is anchored through the magnetic field by the gas
around the cloud center. The magnetic fields in Figures \ref{eigenB002} 
and \ref{eigenB005} are  loosely bent because the roots
are fixed on the outer boundary.  The magnetic tension works against 
fragmentation as expected from the variational principle shown in Appendix A.
When the initial magnetic field is stronger, the density perturbation 
is smaller.   We again note that the unstable mode is normalized so that the 
displacement in the $ z $-direction is $ \xi _z (0,0) = - H $.

\begin{figure}
\plotone{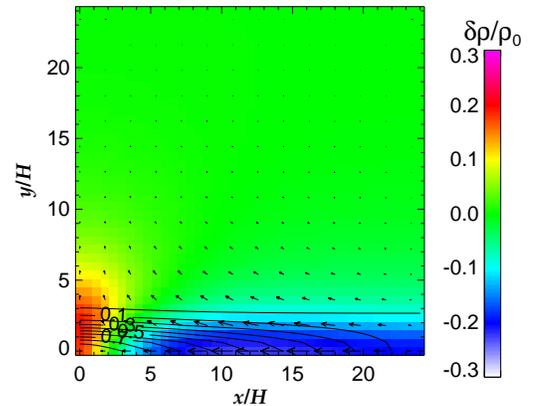}
\caption{Same as Fig.~\ref{eigenB002} except $ \beta = 2 $. 
\label{eigenB100} }
\end{figure}

Figure~\ref{eigenB100} is the same as Figure~\ref{eigenB0} except 
for the equilibrium model,  $ \beta = 2 $.   The growth rate is much smaller
than that for $ B _0 = 0 $.  The displacement is negligibly small in the region $ | y|  > 5 H$, where the 
magnetic flux tube is subcritical.
Note that the displacement is toward the origin along the 
$ x $-axis, while it is away from the origin in the $ y $-direction. 
The $ \delta \varrho > 0 $ region  is concentrated around the $ y $-axis.
This density enhancement is mainly due to the $ x $- and $ z $-components
of the displacement.  The $ y $-component elongates the density enhancement
along the $ y $-axis. 

\begin{figure}
\plotone{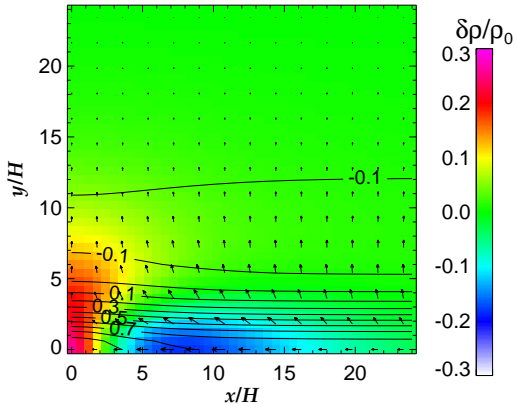}
\caption{Eigenmode for  the free boundary.  The wavenumber
and initial plasma beta are set as $ k H = 0.2 $ and $ \beta = 4 $, respectively.   
The notation
is the same as that of Fig.~\ref{eigenB0}. \label{eigenMFB050} }
\end{figure}

Not only the growth rate but also the eigenfunction depends on
the boundary condition.
Figure~\ref{eigenMFB050} is the same as Figure~\ref{eigenB002}  except the boundary is
free and $ \beta = 4 $.  The $ y $-component
of the displacement has a large value even in the region of $ x \ga 10 H $.
The $ z $-component of the displacement changes its sign around 
$ y \simeq 9 H $.

\begin{figure}
\plotone{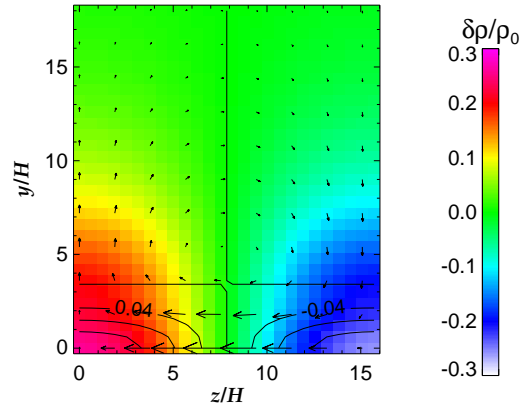}
\caption{The displacement in the $ x = 0 $ plane is shown by the
arrows for the eigenmode shown in Fig.~\ref{eigenMFB050}.  The color indicates
$\delta \rho (0, y, z)/ \rho _0 (0,y,z)$, while the contours are of 
$\delta B _x (0, y, z)/ B _0$. \label{eigenMFB050z} }
\end{figure}

When the free boundary is applied, the magnetic field lines can rotate freely
around the $ x $-axis, as shown in Figure~\ref{eigenMFB050z}.
Figure~\ref{eigenMFB050z} shows the perturbation in the $ x = 0 $
plane for the eigenmode depicted in Figure~\ref{eigenMFB050}.   The arrows
indicate the displacement, while the color indicates the relative change
in the density, $\delta \rho (0, y, z)/\rho _0 (0,y,z)$.   
The contours depict the
change in the $ x $-component of the magnetic field
in intervals of $ \delta B _x (0, y, z)  / B _0 = 0.04 $.  The 
displacement is essentially incompressible circulation, which deforms 
the shape of the filamentary cloud.    The cloud diameter increases in the $y$-direction
at the points where the density increases.  In other words, each dense fragment 
expands in the $ y $-direction.
It should be noted that the cloud changes its
form in the opposite way when the magnetic field is absent or longitudinal.
In that case, the cloud diameter decreases at the points where 
the density increases (see Figure~\ref{eigenB0}). 
The density enhancement is mainly due to the displacement along the density
gradient $ \mbox{\boldmath$\xi$} \cdot \mbox{\boldmath$\nabla$} \rho _0 $.
The compression by the displacement perpendicular to the magnetic field
is very weak.  Compared with the relative change in the density, 
the change in the magnetic field  is much smaller,
i.e., $ \left| \delta B _x/ B _0 \right| \ll | \delta \rho |/ \rho _0 $.

\begin{figure}
\plotone{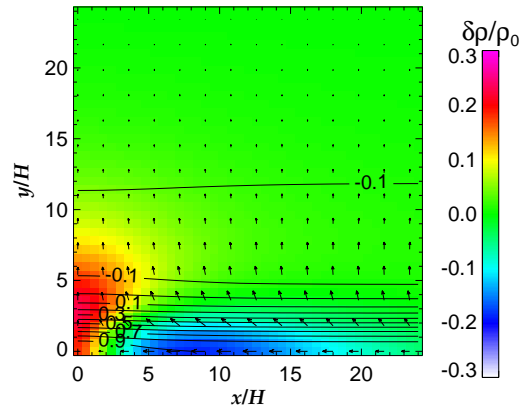}
\caption{Same as Fig.~\ref{eigenMFB050} except $ \beta = 2 $.  
\label{eigenMFB100} }
\end{figure}

Figures~\ref{eigenMFB100} is the same
as Figure~\ref{eigenMFB050} except $ \beta $ = 2.  
The $ z $-component of displacement, $ \xi _z $, changes sign 
around $ y \simeq 12 H $.
In contrast to the case of the fixed boundary, the $y$-component
of displacement is appreciably large in the low-density region.
This is because magnetic tension does not arise by displacement
when the free boundary is applied. 

\begin{figure}
\plotone{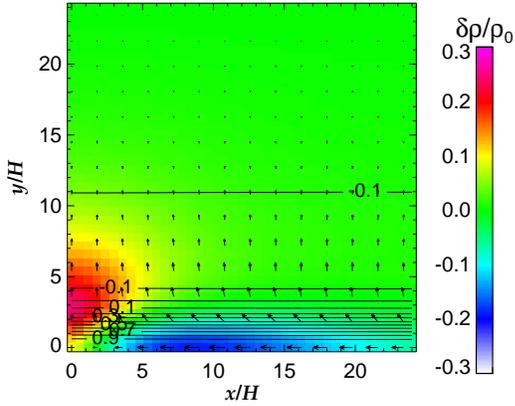}
\caption{Same as Fig.~\ref{eigenMFB050} except 
$\beta = 0.2 $. \label{eigenMFB1000} }
\end{figure}

Figure~\ref{eigenMFB1000} is the same as Figure~\ref{eigenMFB050}
except $ \beta = 0.2 $.   Although the model cloud is subcritical [see
Equation (\ref{betac})], it is unstable.  In other words, it is unstable though
the magnetic
energy dominates over the gravitational energy.    

The free boundary allows rearrangement of the magnetic flux tubes in the
$ yz $ plane while keeping them straight and therefore without induction
of magnetic force.  The rearrangement reduces the gravitational energy
if it gathers more massive tubes, i.e., those tubes initially located near the cloud axis.
The gravitational energy release by the rearrangement depends little
on the magnetic field.  Thus, the growth rate also depends little on  $ B _0 $.
This mode is not taken into account in \citet{hanawa15}, where the
mass-to-flux ratio is assumed to be constant in the stability analysis
for simplicity.

Although it is due to the
self-gravity of the gas, the instability may not result in dynamical collapse,
since the displacement is dominated by circulation.  The change
in gravity is due to the change in the cloud shape.
This instability is similar to that found by \citet{nagai98}.
They found that a magnetized sheet-like cloud is unstable due to
the self-gravity even when the cloud magnetic field is very strong.
We discuss the choice of boundary conditions in the next section.

\section{DISCUSSIONS}

We used a very simplified model for the filamentary cloud in our stability analysis.  The medium
is approximated by an isothermal gas and turbulence is not taken into account.  These simplifications
make the MHD stability analysis feasible.   When the magnetic field is perpendicular to the
cloud axis, we need to distinguish the direction parallel to the field line from that perpendicular
to both the cloud axis and the magnetic field.  Thus, the normal mode analysis cannot be reduced to 
a one-dimensional problem, unlike the previous MHD stability analyses for a filamentary cloud by
\cite{hanawa93,fiege00}.   Our analysis  is formulated as a two-dimensional problem and its 
computational cost is much higher than that for a one-dimensional problem.

Our analysis has shown that the stability depends critically on the boundary assumed for $ x ^2 + y ^2 \rightarrow \infty $.
The dependence comes
from the fact that the filamentary clouds are connected with the outer boundary through the magnetic
field line.   The cloud is stable if the magnetic field is substantial and the field lines are fixed  in a region 
far from it.   On the other hand, the cloud is unstable even  when the magnetic field is extremely strong 
if the field lines are free to move in a region far from it.  This implies that the magnetic field outside
the cloud is important for the stability.    Such a dependence on the outer boundary was
not found in the previous MHD analysis {\citep[e.g.,][]{hanawa93}}, since the magnetic field lines were confined within the cloud.

In this section, we discuss numerical accuracy in \S 4.1 and the effects of magnetic field
orientation in \S 4.2.

\subsection{Numerical Accuracy}

First we examine the accuracy of the growth rate obtained in our numerical
analysis.   The  numerically obtained values should contain some numerical
errors due to the finite spatial resolution, $ \Delta x = \Delta y = 0.6 H $.
The spatial resolution is only moderately high because the full width at half maximum (FWHM)  of the model cloud is $ 3.641 H $.   Despite the moderate
spatial resolution, matrix $ \mbox{\boldmath$A$} $ has dimension
\begin{eqnarray}
n_A & = & 3 n _x n _y + 2 n _x + 2 n _y + 1= 4961, 
\end{eqnarray}
so solving the eigenvalue problem takes considerable computation time.
The computation time is roughly proportional to $ n _x ^3 n _y ^3 $ since
the computation time increases in proportion to $ n _A ^3 $.
 Hence, it is not easy to achieve higher spatial resolution to reduce 
 the numerical error in our 2D analysis.
 
 We examine the case of $ B _0 = 0 $ to assess the error due to
 the finite spatial resolution.  When $ B _0 = 0 $, the equilibrium 
 model is symmetric around the axis and therefore the perturbation is also
 symmetric.  The eigenvalue problem reduces to a 1D problem, as
 shown in Appendix B.   We can achieve very high spatial 
 resolution to eliminate the numerical error due to the finite spatial
 resolution.
 
 As shown in Appendix \ref{noB0}, our numerical method has second-order
 accuracy in space.  The numerical error is proportional to the square of
 the spatial resolution, $ \Delta r $, in our 1D analysis and is evaluated as
 \begin{eqnarray}
 \frac{\Delta (\sigma ^2)}{4 \pi G \rho _c}
  \simeq 3.4 \times 10 ^{-4} \left( \frac{\Delta r}{0.1 H} \right) ^2 ,
 \end{eqnarray}
 from Figure~\ref{C1D2D} for the mode of $  kH = 0.25 $.

Table~\ref{1D} compares the growth rate squared, $ \sigma ^2 $, obtained
in our 1D and 2D analyses.    The spatial resolutions are $ \Delta r  = 0.1 H $
and $ \Delta x = \Delta y = 0.6 H $ in the 1D and 2D analyses, respectively.
Despite the large difference in the formal spatial resolutions, the difference
in the obtained growth rates is as small as
\begin{eqnarray}
-1.1 \times 10 ^{-3} < \frac{\left( \sigma ^2 \right) _{\rm 2D} - \left( \sigma ^2 \right) _{\rm 1D}}
{4 \pi G \rho _c} < -3 \times 10 ^{-4} .
\end{eqnarray}
This difference is comparable to the expected numerical error for the growth rate obtained in our
1D analysis.    Thus, we estimate that the numerical error of our 2D analysis is
about $ \Delta \left( \sigma ^2 \right) \simeq 10 ^{-3} (4 \pi G \rho _0) $. 
Although the formal spatial resolution is $ \Delta x = \Delta y = 0.6 H $,
the effective spatial resolution is likely to be higher since the displacement
is evaluated not only on the $ x $- and $ y $-axes but at  various points
in the $xy$ plane.   

The accuracy mentioned above is consistent with the spatial extent of
the displacement. As shown in Figure~\ref{eigenB0}, the displacement
is large in the region $ \sqrt{x ^2 + y ^2} \la 10 H $, which is covered
by $ 16 \times 16 $ grid points.   The error due to the finite difference
is estimated to be $ \approx [1/(2\times16)] ^{-2} \simeq 10 ^{-3} $, since
the difference equations are of second-order accuracy.
The factor 2 is introduced to take account of the region of negative
$ x $ and $ y $.  

The presence of the magnetic field may increase the numerical error since the magnetic
field confines the displacement in an especially narrow region when the 
fixed boundary condition is applied.  Nevertheless, the error is likely to be moderate
since the eigenmode is well resolved, as shown in the previous section.
We expect that the numerical error is smaller than 
$ \Delta ( \sigma ^2 ) < 5 \times 10 ^{-3} (4 \pi G \rho _c ) $ and
does not have serious effects on our analysis.

\subsection{Effects of Magnetic Field Perpendicular to the Cloud}

Next we discuss the dependence of the instability on the direction of 
the magnetic field.  As shown in the previous section, the magnetic field
works against fragmentation and reduces the growth rate of
instability  when it is perpendicular to the cloud axis.  On the other hand, 
the magnetic field does not suppress the instability when it is parallel to
the cloud axis \citep[see, e.g.,][]{stodolkiewicz63,hanawa93}.  
A toroidal or helical magnetic field stabilizes fragmentation of the
cloud and reduces the growth rate of the sausage (axisymmetric) 
mode \citep{fiege00}, although it induces the kink (non-axisymmetric) mode.
These modes are characterized by the geometry of the magnetic field, namely,
whether the magnetic field is bent or compressed by the fragmentation.

Before applying our analysis to observed filamentary molecular clouds,
we need to discuss the boundary conditions employed in our analysis.
As shown in the previous section, the boundary conditions greatly affect the instability.
When the free boundary condition is applied, the filamentary cloud is
unstable even when the magnetic field is extremely strong.
In other words, no magnetic field can suppress the instability
if the displacement does not vanish in the region very far from the
cloud.  The unstable mode exists because the density decreases
in proportion to $ \rho _0 \propto r ^{-4}$, where
$ r $ denotes the radial distance from the cloud axis. 
In our model, the displacement travels from the
cloud center to infinity through the Alfv\'en wave in a finite time,
\begin{eqnarray}
\tau _{\rm A} & = & \int _0 ^\infty \frac{\sqrt{4 \pi \rho_0}}{B _0} dx  = 
\frac{2 \pi \sqrt{2 \pi \rho _c} H}{B _0} .
\end{eqnarray} 
The Alfv\'en transit time may be much longer in
reality because the density is higher or the magnetic field is 
weaker in the region far from the cloud.   
\citet{palmeirim13} reported that the mean radial column density fits
a Plummer-like density distribution,
\begin{eqnarray}
\rho _{\rm P} & = & \frac{\rho _c}{\left[ 1 + \left( r / R _{\rm flat} ^2 \right)
\right] ^{p/2} } ,
\end{eqnarray}
where $ R _{\rm flat} $ denotes the radius of the inner flat region.
In the best-fit model, $ R _{\rm flat} = 0.07 \pm 0.02~\mbox{pc} $
and $  p = 2.0 \pm 0.4 $ for the B211 filament.
The density slope is significantly shallower than that for the isothermal
equilibrium, $ p =  4 $.  
Our model cloud is highly idealized, especially in the region very far
from the cloud axis.  The shallower slope might  be due to 
turbulence, which is likely to be larger in a lower density gas.
We again note that the striations running perpendicular
to the main filament \citep[see, e.g.,][]{sugitani11,palmeirim13} 
are unlikely to move
easily by magnetic force since the Alfv\'en velocity is lowered by the
relatively high density.  If the Alfv\'en transit timescale  is longer than the
free-fall timescale, the fixed boundary condition may be appropriate.
We need to account for the  low-density region surrounding
 the filamentary cloud when applying our analysis.

\cite{tomisaka14} made a model for a filamentary cloud in which the magnetic
field lines lie in the plane perpendicular to the cloud axis.   In his equilibrium 
model, the cloud is supported in part by the magnetic tension.   The magnetic
field is weaker at a greater distance from the cloud axis. 
In other studies with 3D MHD simulations, the magnetic fields are not
uniform around the filamentary clouds formed 
\citep[see, e.g.,][]{nakamura08,klassen17}.   
If the density is very low in the periphery of the filamentary cloud, the
magnetic field should be quasi force-free and thus nearly straight.
It is also well known that the magnetic field is stronger where the
column density is higher \citep[see, e.g.,][]{li14}.

Our result for the free boundary is quite similar to those obtained 
by \citet{nagai98} and \citet{fiege00}, who studied the stability of an isothermal 
gas layer and a filament confined by the outer gas pressure, respectively. 

\citet{nagai98} assumed that the magnetic field is uniform and parallel to
the gas midplane.   This gas layer suffers from two types of
instability: fragmentation due to displacement along the magnetic field 
and distortion of the surface layer.   When the gas layer is thick and
consequently mostly gravitationally bound,  the instability due to displacement dominates.
When the gas layer is thin and consequently pressure bound, the distortion dominates and
the gas layer fragments into filaments.   The magnetic fields remain
straight as the instability due to distortion develops.  
As a result, the magnetized gas sheet is unstable and the
growth rate remains constant in the limit of a strong magnetic field,
as in the case of the free boundary condition in our stability analysis.
Although the latter instability is due
to the self-gravity, the filaments formed do not collapse directly, 
since the filaments expand in the direction normal to the initial sheet by the instability.
It should again be noted that this instability develops even when the
sheet is extremely thin and the self-gravity is not important.
It seems that the this instability
is analogous to the instability for the free boundary.   

\citet{fiege00} showed that their truncated Ostriker model is unstable
against fragmentation.   The model filamentary cloud is assumed to be
surrounded by hot gas of negligibly small density.  
Their initial model was similar to that of \cite{nagai98} except
for the geometry and the magnetic field.
When the pressure of the hot gas is comparable to that at the filament center,
circulation dominates in the unstable mode, as in our model for
the free boundary (see their Figure 4).  Each dense clump expands in the
radial direction as a result of the instability.
 The change in the gravitational
potential is mainly due to that in the boundary between hot and cold
gases (i.e., the boundary of the dense gas), as in our free boundary model.
It seems that a relatively strong magnetic field increases the effective temperature 
of the gas through magnetic pressure and therefore we obtain similar results. 

As shown in \S 2, the model cloud is subcritical when the plasma beta is
as low as 0.405.  It is well known that molecular clouds collapse 
dynamically only when the cloud is supercritical.  Thus, the subcritical
cloud forms clumps supported in part by the magnetic force, 
not by dynamically collapsing cores.  In other words, fragmentation does
not result in direct  star formation.   

The above argument implies that subcritical clumps can be formed
via fragmentation of a filamentary cloud if the free boundary is the case.
A subcritical filamentary cloud may suffers from fragmentation instability. 
The instability is not likely to result in the dynamical collapse  to form 
gravitationally bound clumps supported mainly by magnetic field.  Such 
clumps may form stars  through quasi-static contraction,
i.e., after substantial amount of magnetic field is liberated through the 
ambipolar diffusion.  

Our analysis suggests to study low density region
surrounding a filamentary cloud.
Motion in the high-density region
may propagate into the outer region through the Alfv\'en wave, 
when the filamentary cloud fragments.
It would be meaningful to observe the velocity field around the filamentary
structure.  The motion perpendicular to the magnetic field may 
contain information on the motion in the dense region.
The Alfv\'en speed is proportional to the inverse square root of the density
($ \propto \rho ^{-1/2} $).  Thus the motion is faster when  the density is lower. 
The density distribution may be controlled by turbulence.   If it is the case,
turbulence may affect the stability of a filamentary cloud indirectly through
the boundary condition.

\acknowledgements

We thank an anonymous referee for useful comments to clarify the arguments.  This work
was supported by JSPS KAKENHI Grant Number JP15K05032.

\software{LAPACK, Linear Package Algebra \citep{anderson99}}

\appendix

\section{Variational Principle}

In this appendix, we prove that the growth rate is either real or pure imaginary.
For this purpose, we integrate the inner product of Equation (\ref{motion}) and
$ \mbox{\boldmath$\xi$}^* $ over the volume, where the asterisk denotes the
complex conjugate.   The integral is expressed as
\begin{eqnarray}
\sigma ^2 I & = & W _{\rm T} + W _{\rm G} + W _{\rm M} , \\
I & = & \int _{-\infty} ^{+\infty} \! \! \! \int _{-\infty} ^{+\infty} \! \! \!  \int _0 ^{2\pi/k} 
\rho | \mbox{\boldmath$\xi$} | ^2 dx dy dz, \\
W _{\rm T} & = & - c _s ^2 \int _{-\infty} ^{+\infty} \! \! \! \int _{-\infty} ^{+\infty} \! \! \!  \int _0 ^{2\pi/k}
 \rho _0 \mbox{\boldmath$\xi$} ^* \cdot 
 \mbox{\boldmath$\nabla$} \left( \frac{\delta \rho}{\rho _0} \right) dxdydz , \\
W _{\rm G} & = & - \int _{-\infty} ^{+\infty} \! \! \! \int _{-\infty} ^{+\infty} \! \! \!  \int _0 ^{2\pi/k}
 \rho _0 \mbox{\boldmath$\xi$} ^* \cdot 
\mbox{\boldmath$\nabla$} \delta \psi dxdydz , \\
W _{\rm M} & = & \int _{-\infty} ^{+\infty} \! \! \! \int _{-\infty} ^{+\infty} \! \! \!  \int _0 ^{2\pi/k}
\mbox{\boldmath$\xi$} ^* \cdot 
\left( \delta \mbox{\boldmath$J$} \times \mbox{\boldmath$B$} _0 \right)  dxdydz ,
\end{eqnarray}
The integral $ I $ is real and positive for any $ \mbox{\boldmath$\xi$} $.  The integral
$ W _{\rm T} $ is also proved to be real and negative, since
\begin{eqnarray}
W _{\rm T} & = &  c _s ^2 \int _{-\infty} ^{+\infty} \! \! \! \int _{-\infty} ^{+\infty} \! \! \!  \int _0 ^{2\pi/k}
\mbox{\boldmath$\nabla$} \cdot \left( \rho _0 \mbox{\boldmath$\xi$} ^* \right) 
\left( \frac{\delta \rho}{\rho _0} \right) dxdydz   \\
& = & - c _s ^2 \int _{-\infty} ^{+\infty} \! \! \! \int _{-\infty} ^{+\infty} \! \! \!  \int _0 ^{2\pi/k}
\frac{|\delta \rho  | ^2}{\rho _0} dxdydz .
 \end{eqnarray}
 Here we use the boundary condition at infinity and Equation (\ref{continuity}) in the integration
 by parts.  Similarly, the integral $ W _{\rm G} $ is proved to be real and positive for any
  perturbation, since
\begin{eqnarray}
W _{\rm G} & = &  \int _{-\infty} ^{+\infty} \! \! \! \int _{-\infty} ^{+\infty} \! \! \!  \int _0 ^{2\pi/k}
\mbox{\boldmath$\nabla$} \cdot \left(  \rho _0 \mbox{\boldmath$\xi$} ^* \right)
\delta \psi dxdydz \\
& = & -  \int _{-\infty} ^{+\infty} \! \! \! \int _{-\infty} ^{+\infty} \! \! \!  \int _0 ^{2\pi/k}
\delta \rho ^* \delta \psi dxdydz \\
& = & -  \frac{1}{4\pi G} \int _{-\infty} ^{+\infty} \! \! \! \int _{-\infty} ^{+\infty} \! \! \!  \int _0 ^{2\pi/k}
\Delta \delta \psi ^* \delta \psi dxdydz  \\
& = & \frac{1}{4\pi G} \int _{-\infty} ^{+\infty} \! \! \! \int _{-\infty} ^{+\infty} \! \! \!  \int _0 ^{2\pi/k}
\left| \mbox{\boldmath$\nabla$} \delta \psi \right| ^2  dxdydz .
\end{eqnarray}
The integral $ W _{\rm M} $ is proved to be real and negative, since
\begin{eqnarray}
W _{\rm M} & = & - \int _{-\infty} ^{+\infty} \! \! \! \int _{-\infty} ^{+\infty} \! \! \!  \int _0 ^{2\pi/k}
\delta \mbox{\boldmath$J$} \cdot 
\left( \delta \mbox{\boldmath$\xi$} ^* \times \mbox{\boldmath$B$} _0 \right)  dxdydz \\
& = & - \frac{1}{4\pi} \int _{-\infty} ^{+\infty} \! \! \! \int _{-\infty} ^{+\infty} \! \! \!  \int _0 ^{2\pi/k}
\mbox{\boldmath$\nabla$} \times \delta \mbox{\boldmath$B$} \cdot 
\left( \delta \mbox{\boldmath$\xi$} ^* \times \mbox{\boldmath$B$} _0 \right)  dxdydz \\
& =  & - \frac{1}{4\pi} \int _{-\infty} ^{+\infty} \! \! \! \int _{-\infty} ^{+\infty} \! \! \!  \int _0 ^{2\pi/k}
\delta \mbox{\boldmath$B$} \cdot 
\mbox{\boldmath$\nabla$} \times \left( \delta \mbox{\boldmath$\xi$} ^* \times \mbox{\boldmath$B$} _0 \right)  
dxdydz \\
& = & - \frac{1}{4\pi} \int _{-\infty} ^{+\infty} \! \! \! \int _{-\infty} ^{+\infty} \! \! \!  \int _0 ^{2\pi/k}
 \left| \delta \mbox{\boldmath$B$} \right| ^2  dxdydz .
\end{eqnarray}
Either $ \mbox{\boldmath$\xi$}$ or $ \delta \mbox{\boldmath$B$} $ is assumed to vanish at infinity.
$ \mbox{\boldmath$\xi$}$  corresponds to the fixed boundary, whereas $ \delta \mbox{\boldmath$B$} $ corresponds to the free boundary.  
Thus, the square of the growth rate, $ \sigma^2 $, should be real since
\begin{eqnarray}
\sigma ^2 & = & \frac{W _{\rm T} + W _{\rm G} + W _{\rm M}}{I} . \label{variationalP}
\end{eqnarray}
It is also clear from Equation (\ref{variationalP}) that the filament can be unstable only through 
the self-gravity, $ W _{\rm G} $.

\section{Matrix Elements} \label{matrix-elements}

This appendix describes each component of the vector $ \mbox{\boldmath$\zeta$} $,
and matrixes  $ \mbox{\boldmath$A$} $, $ \mbox{\boldmath$B$} $, and $ \mbox{\boldmath$C$} $.

The $m$-th element of the vector $ \mbox{\boldmath$\zeta$} $  is set so that
\begin{eqnarray}
\zeta _{m} & = & \xi _{x,i-1/2,j} , \hskip 10pt (i=1, 2, \dots , n_x~\mbox{and}~j=0, 1, \dots, n _y) \\
m & = & i + n _x j , \label{mx}
\end{eqnarray}
for $ 1 \le m \le n _x \left( n _y + 1 \right)$.  Indexes $ i $ and $ j $ are derived from $ m $ by
\begin{eqnarray}
i & = & [ ( m - 1) \, (\mbox{mod} \,  n _x) ] + 1 , \\
j & = & \frac{m - i}{n _x} .
\end{eqnarray}
Other elements of the vector, $\mbox{\boldmath$\zeta$} $,  are set as 
\begin{eqnarray}
\zeta _m & = & \xi _{y,i,j-1/2} ,  \hskip 10pt (i=0, 1, \dots , n_x~\mbox{and}~j = 1, 2, \dots, n _y)\\
m & = & i + 1 + \left( n _x + 1 \right) \left(j - 1 \right) + n _x \left( n _y + 1 \right) , \label{my} \\
i & = & \left\{ \left[ m - n _x \left( n _y + 1 \right) - 1 \right] \, \left( \mbox{mod} \, n _x + 1 \right) \right\} + 1 \\
j & = & \frac{m  - i - n _x \left( n _y + 1 \right)}{n _x + 1} , 
\end{eqnarray}
for $  n _x  \left(n _y +  1\right)  < m \le 2 n_x n_y +  n _x +  n _y  $, and
\begin{eqnarray}
\zeta _m & = & \xi _{z,i,j} ,  \hskip 10pt (i=0, 1, \dots , n_x~\mbox{and}~j=0, 1, \dots, n _y)\\
m & = & i + 1 + \left( n _x + 1 \right) j + 2 n _x  n _y + n _x +  n _y, \label{mz}
\end{eqnarray}
for $ 2 n _x n _y + n _x + n _y  < m \le 3 n _x n _y + 2 n _x + 2 n _y  + 1 $.  
The changes in the density, current density, and gravity are evaluated once the displacement,
$ \mbox{\boldmath$\zeta$} $, is given.  All the forces are  proportional to the displacement
$ \mbox{\boldmath$\zeta$} $, and the magnetic force is also proportional to $ B _0 ^2 $. 
Hence, Equations (\ref{motion1}) through (\ref{motion3}) are transformed into Equation (\ref{algebraic}).

The $ m $-th column of matrix $ \mbox{\boldmath$A$} $ is evaluated by the
following procedure.  First we compute $ \delta \varrho _{i,j} $ for all pairs of
$ i $ and $ j $ according to Equation (\ref{continuity3}) for  $ \zeta _m = 1 $ and $ \zeta _{m^\prime} = 0 $ 
for $ m ^\prime \ne m$.  Note that the change in the density, $ \delta \varrho _{i,j} $,
vanishes for the given displacement except at a few points.  Then the change in the
gravitational potential is evaluated according to Equation (\ref{poisson3}).  The Green's function is defined
as  the solution of
\begin{eqnarray}
\frac{G _{i+1,j,i^\prime,j^\prime} - G _{i-1,j,i^\prime,j^\prime}}{\Delta x ^2} +
\frac{G _{i,j+1,i^\prime,j^\prime} - G _{i,j-1,i^\prime,j^\prime}}{\Delta y ^2} 
- \left( \frac{2}{\Delta x^2} + \frac{2}{\Delta y ^2} + k ^2 \right) G _{i,j,i^\prime,j^\prime}
& = & 4 \pi G \delta _{i,i^\prime} \delta _{j,j^\prime} ,
\end{eqnarray}
where $ \delta  _{i,i^\prime}$ and $ \delta _{j,j^\prime} $ denote the Kronecker's delta.
The solution should satisfy the boundary condition on the $ y $-axis ($ x = 0 $),
\begin{eqnarray}
G _{-1,j,i^\prime,j^\prime} & = & G _{1,j,i^\prime,j^\prime} ,
\end{eqnarray}
and that on the $ x $-axis ($ y = 0 $),
\begin{eqnarray}
G _{i,-1,i^\prime,j^\prime} & = & G _{i,1,i^\prime,j^\prime} ,
\end{eqnarray}
for any  $ i $, $ j $, $ i ^\prime $, and $ j^\prime $.   The boundary conditions are taken
into account by the mirror image method: we superimpose the solutions for the mirror
images in which the
boundary conditions are set as $ G _{i,j,i^\prime, j^\prime} = 0 $ in the
region $ | i - i ^\prime | \Delta x \gg H $ or $ | j - j ^\prime | \Delta y \gg H $.  We use 
Gauss-Seidel iteration to obtain the Green's function with the boundary condition set
as $ | i - i ^\prime | = 2 n _x $ and $ | j - j ^\prime | = 2 n _y $.
Using $ \delta \varrho _{i,j} $ and $ \delta \psi _{i,j} $, we evaluate the pressure force
and gravity in the right-hand side of Equations (\ref{motion1}) through (\ref{motion3}) 
to obtain the $ m $-th column of matrix $\mbox{\boldmath$A$} $.   Equations (\ref{motion1}),
(\ref{motion2}), and (\ref{motion3}) provide the first $ n _x \left( n _y + 1 \right) $, second
$ \left( n _x + 1 \right) n _y $, and last $ (n _x + 1) (n _y + 1) $ components of
the $ m $-th column of $ \mbox{\boldmath$A$} $.  The row number is evaluated from $ i $ and $ j $ according to
Equations  (\ref{mx}), (\ref{my}), and (\ref{mz}). 

Matrix  $ \mbox{\boldmath$B$} $ is diagonal with diagonal elements representing the equilibrium density, 
$ \rho _0 $, at the point where the displacement is evaluated.

Matrix $ \mbox{\boldmath$C$} $ is obtained in a similar manner.  The $ m $-th column of $ \mbox{\boldmath$C$} $
denotes the magnetic force when $ \zeta _{m^\prime} = \delta _{m,m^\prime} $ and $ B _0 = \sqrt{4\pi \rho _c} c _s $.

\section{Case of No Magnetic Field} \label{noB0}

In this appendix, we examine the instability of an unmagnetized
filamentary cloud.  When $ \mbox{\boldmath$B$} _0  = 0 $, 
the equilibrium model is symmetric around the axis.  Hence, the 
unstable mode is also symmetric around the
axis.   Then, density, displacement, and potential can be described
as
\begin{eqnarray}
\rho & = & \rho _0 + \delta \varrho (r) \cos k z , \\
\mbox{\boldmath$\xi$} & = & \xi _r (r) \cos k z \mbox{\boldmath$e$} _r + \xi _z (r)
\sin k z \mbox{\boldmath$e$} _z , \\
\psi & = & \psi _0 + \delta \psi (r) \cos k z ,
\end{eqnarray}
in cylindrical coordinates $ (r, \theta, z ) $. The perturbation equations are
written as
\begin{eqnarray}
\delta \varrho & = & - \frac{1}{r} \frac{\partial}{\partial r} \left( r \rho _0 \xi _r \right)
- k \rho _0 \xi _z , \label{noMrho} \\
\sigma ^2 \rho _0 \xi _r & = & - c _s ^2 \rho _0 \frac{\partial}{\partial r} \left( \frac{\delta \varrho}{\rho _0} \right)
- \rho _0  \frac{\partial}{\partial r} \delta \psi , \label{noMxir} , \\
\sigma ^2 \rho _0 \xi _z & = & k c _s ^2 \delta \varrho + k \rho _0 \delta \psi  , \label{noMxiz}  \\
4 \pi G \delta \varrho & = & \frac{1}{r} \frac{\partial}{\partial r} \left( r \frac{\partial}{\partial r} \delta \psi \right) 
- k ^2 \delta \psi .
\end{eqnarray}
From 
Equations (\ref{noMxir}) and (\ref{noMxiz}), we  obtain
\begin{eqnarray}
\frac{\partial \xi _z}{\partial r} + k \xi _r & = & 0 . \label{rotxi}
\end{eqnarray}

We solve the perturbation equations numerically by the following procedure.
First we discretize Equation (\ref{rotxi}) in the form 
\begin{eqnarray}
\xi _{r,j+1/2} & = & - \frac{1}{k \Delta r} \left( \xi _{z,j+1} - \xi _{z,j} \right) , \label{xir} 
\end{eqnarray}
where $ \xi _{r,j+1/2} $ and $ \xi _{z,j} $ denote the values at $ r =  r _{j+1/2} 
=  (j+1/2) \Delta r $
and $ r = r _j =  j \Delta r $, respectively.  The change in density is evaluated as
\begin{eqnarray}
\delta \varrho _j & = & - \frac{1}{r _j \Delta r} \left( r _{j+1/2} \rho _{0,j+1/2} \xi _{r,j+1/2} 
- r _{j-1/2} \rho _{0,j-1/2} \xi _{r,j-1/2} \right)
- k \rho _{0,j} \xi _{z,j} 
\end{eqnarray}
by discretizing Equation (\ref{noMrho}), where $ \delta \varrho _{j} $ and 
$ \rho _{0,j+1/2} $ denote the values at  $ r = r _j $ and $ r _{j+1/2}$, respectively. 
The change in the gravitational potential, $\delta \psi $, is obtained by solving 
the discretized Poisson equation
\begin{eqnarray}
4 \pi G \delta \varrho _{j} & = & 
\left\{ 
\begin{array}{ll}
\displaystyle 2 \frac{\delta \psi _1 - \delta \psi _0}{\Delta r ^2}
- k ^2 \delta \psi _0  & \hskip 20pt ( j = 0) \\
\displaystyle - \frac{ r  _{j+1/2} \delta \psi _{j+1} - 2 r _j 
\delta \psi _j + r _{j-1/2} \delta \psi _{j-1}}{r _j \Delta r ^2}
- k ^2 \delta \psi _j  & \hskip 20pt (j=1, 2, \dots , n)
\end{array}  \right. ,
\end{eqnarray}
with the boundary condition
\begin{eqnarray}
\delta \psi _{n+1} = \left( 1 + k \Delta r \right) ^{-1} \delta \psi _{n} , 
\label{psi-boundary}
\end{eqnarray}
where $ j = n $ denotes the outermost grid point while $ j = n + 1$ does
the one adjacent outside.

Using Equations (\ref{xir}) through (\ref{psi-boundary}), we derive $ \xi _{r,j-1/2} $, 
$ \delta \varrho _j $, and $ \delta \psi _j $ for a given set of $ \xi _{z,j} $. 
Then we can evaluate the right-hand side of Equation (\ref{noMxiz}) in the form of
\begin{eqnarray}
\sigma ^2 \rho _{0,j} \xi _{z,j} & = & \sum _{i} F _{ji} \xi _{z,i} .
\end{eqnarray}
The growth rate $ \sigma $ is numerically obtained by using
the Linear Algebra Package, LAPACK \citep{anderson99}).

\begin{figure}[h]
\epsscale{0.5}
\plotone{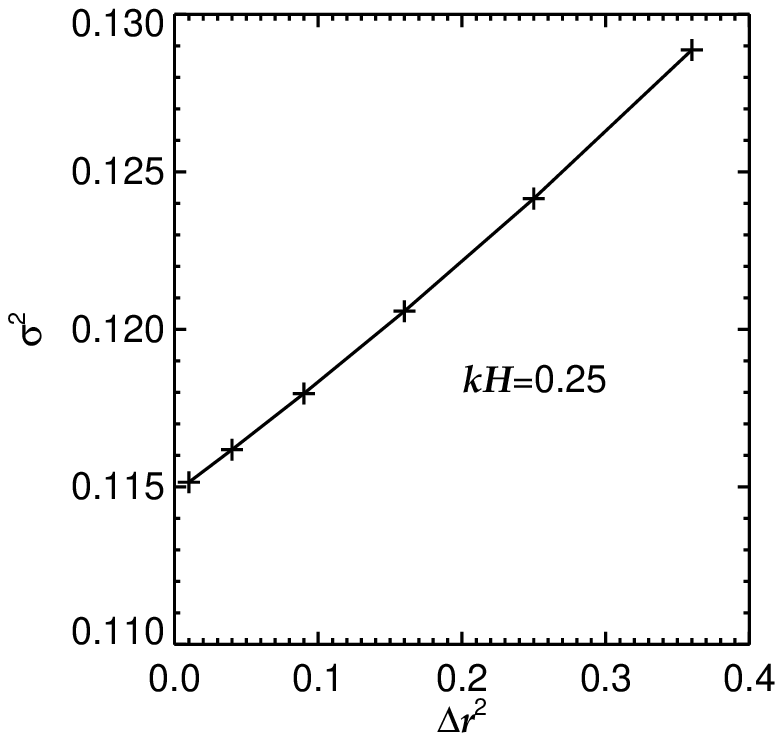}
\caption{Dependence of the growth rate squared on the
spatial resolution for $ B _0 = 0 $ and $ k H = 0.25 $. \label{1D}}
\end{figure}

In this 1D stability analysis, we can evaluate the numerical error due to the
discretization.  Figure \ref{1D} shows $ \sigma ^2 $ as a function of the
spatial resolution $ \Delta r $ for the mode $ k H = 0.25 $.   The outer boundary
is placed at $ r = 60 H $, which is far away enough and does not affect the result.  
The discretization is of second-order accuracy and the growth rate 
is expressed as
\begin{eqnarray}
\sigma ^2 (\Delta r; k H =0.25) & = & 4 \pi G \rho _c \left[ 
0.114803 + 3.44 \times 10 ^{-4} \left( \frac{\Delta r} 
{0.1 H} \right) ^2 \right] . \end{eqnarray} 
We conclude that the growth rate is highly accurate when
$ \Delta r = 0.1~H $ in the1D analysis.  

Table \ref{C1D2D} compares the growth rates obtained in our 2D analysis with
the growth rates in our 1D analysis for various $ kH$. The middle column lists $ \sigma ^2 $ values obtained
in the 1D analysis, while the right column lists those obtained in the 2D analysis
for $ B _0 = 0 $.
The spatial resolution is $ \Delta r  = 0.1~H $ in the 1D analysis and $ \Delta x = 
\Delta y = 0.6~kH $ in the 2D analysis.   The outer boundary is set as $ r = 60~H $
in the 1D and $ x ~\mbox{or}~y = 24.3~H $ in the 2D analysis.  
Since the difference is small, we conclude that our 2D analysis gives 
a reliable growth rate at least for $ B _0 = 0 $.

\begin{table}[h]
\caption{Comparison of Growth Rates Squared, $\sigma ^2$, 
Obtained in the 1D and 2D Analyses. \label{C1D2D}}
\begin{center}
\begin{tabular}{lcc}
\hline
$k H $ & 1D  & 2D  \\
\hline
0.05 &  0.02962 & 0.02995 \\
0.10 &  0.06135 & 0.06186 \\
0.15 &  0.08720 & 0.08785 \\
0.20 &  0.10528 & 0.10606 \\
0.25 &  0.11515 & 0.11606 \\
0.30 &  0.11684 & 0.11787 \\
0.35 &  0.11059 & 0.11173 \\
0.40 &  0.09666 & 0.09790 \\
0.45 &  0.07534 & 0.07667 \\
0.50 &  0.04690 & 0.04831\\
\hline
\end{tabular}
\end{center}
\end{table}




\clearpage

\allauthors

\listofchanges

\end{document}